\documentclass{ws-procs975x65}

\def\beq{\begin{equation}}
\def\eeq{\end{equation}}

\begin{document}

\title{ULXs as magnetized sub-Eddington advective accretion flows around stellar mass black holes}
\author{Banibrata Mukhopadhyay}

\address{Department of Physics, Indian Institute of Science,\\
Bangalore 560012, India\\
E-mail: bm@iisc.ac.in\\
www.iisc.ac.in}

%

\begin{abstract}
Ultra-luminous X-ray sources (ULXs) have been puzzling us with a debate whether they consist of an intermediate mass black hole or super-Eddington accretion by a stellar mass black hole. Here we suggest that in the presence of large scale strong magnetic fields and non-negligible vertical motion, the luminosity of ULXs, particularly in their hard states, can be explained with sub-Eddington accretion by stellar mass black holes. In this framework of 2.5D magnetized advective accretion flows, magnetic tension plays the role of transporting matter (equivalent to viscous shear via turbulent viscosity) and we neither require to invoke an intermediate mass black hole nor super-Eddington accretion. Our model explains the sources, like, NGC 1365 X1/X2, M82 X42.3+59, M99 X1 etc. which are in their hard power-law dominated states.
\end{abstract}


\bodymatter


\section{Introduction}

While the existence of stellar mass black holes of mass $M\lesssim 80 M_\odot$ and supermassive black holes of 
mass $M\gtrsim 10^6 M_\odot$ are confirmed, there is no direct evidence for black holes of mass in between.
Scientists believing in the continuous mass distribution argue for the existence of such black holes, called
intermediate mass black hole. However, many others argue that there is no such obvious expectation as 
the origins of stellar mass and intermediate mass black holes are completely different. Nevertheless,
there are ultra-luminous X-ray sources (ULXs) observed in galaxies around, which apparently cannot be explained by the conventional idea
of stellar mass black holes accreting at a sub-Eddington limit. Hence, the proposal is that the sources 
harbor an intermediate mass black hole, particularly when they reveal lower temperature in the underlying
multicolor black hole spectra \cite{lowt}. However, there is another idea behind ULXs that they are 
stellar mass black holes only but accreting at a super-Eddington rate: candidates for slim accretion
disk \cite{abra88}. 

Nevertheless, none of the above ideas is a conventional one. There are significant evidences that X-ray 
binaries are sub-Eddington accretors and there is no direct evidence yet of galactic black hole mass
$M \gtrsim 100M_\odot$ (though the detection of gravitational wave argues for the black hole mass larger
than that determined in X-ray astronomy). Here our story lines start. We show that ULXs in hard states
can be explained by a stellar mass black hole accreting at a sub-Eddington rate with advection in the 
presence of large scale strong magnetic field. Hence, by the interplay between magnetic field and
advection, X-ray binaries could be quite luminous in the hard state. For that we neither require 
an intermediate mass black hole nor super-Eddington accretion. Hence, while still the existence of
intermediate black hole, even appeared as ULX, is not ruled out, some ULXs in hard states, e.g.
NGC 1365 X1/X2, M82 X42.3+59, M99 X1 etc., are suggested to be highly magnetized stellar mass black hole sources only.

We model a combined disk-outflow coupled system with the inclusion of vertical velocity 
and large scale magnetic stress explicitly. This is essentially a 2.5D magnetized advective
accretion disk model. We show that energetics and luminosities of such a flow are in accordance 
with ULXs.

\section{Magnetized disk-outflow coupled system}

We consider a magnetized, viscous, advective disk-outflow/jet symbiotic system 
with cooling around black holes. 
We consider the large scale magnetic and turbulent viscous stresses both and depending on the field strength
one of them may dominate over other.
Here we assume a steady and axisymmetric flow and all the flow parameters: radial velocity $(v_{r})$, specific angular momentum $(\lambda)$, outflow or vertical velocity $(v_{z})$, fluid pressure $(p)$, mass density $(\rho)$, radial $(B_{r})$, azimuthal $(B_{\phi})$, and vertical $(B_{z})$ components of magnetic field, are functions of both radial and vertical coordinates. Throughout we express length variables in units of $GM_{BH}/c^{2}$, where $G$ is the Newton's gravitational constant, $M_{BH}$ the mass of BH, and $c$ the speed of light. Accordingly, we also express 
other variables. 
Hence, the continuity and momentum balance equations are respectively
\begin{align}
&\nabla.\left(r\rho\mathbf{v}\right)=0,\,\,{\rm and}\,\,
(\mathbf{v}.\nabla)\mathbf{v}=\mathbf{F}-\frac{1}{\rho}\nabla \left(p+\frac{B^{2}}{8\pi}\right)+\frac{(\mathbf{B}.\mathbf{\nabla})\mathbf{B}}{4\pi \rho}+\frac{1}{\rho}\nabla .\mathbf{W},
\end{align}
where $\mathbf{v}$ and $\mathbf{B}$ are velocity and magnetic field vectors respectively,
$|\mathbf{F}|$ is the magnitude of the gravitational force for a BH in the pseudo-Newtonian framework \cite{m02}. The importance of generalized viscous shearing stress tensor $(\mathbf{W}=W_{ij})$ is taking care explicitly in this formalism.
Various components of $W_{ij}$ are written in terms of $\alpha$-prescription \cite{ss73} with appropriate modifications 
\cite{mm19}.
We also have to supplement the above equations with the equations for no magnetic monopole and induction, as respectively
\begin{equation}
\nabla .\mathbf{B}=0  \ \text{and} \ \nabla \times \left(\mathbf{v}\times\mathbf{B} \right)+\nu_{m}\nabla^{2}\mathbf{B}=0,
\label{inde}
\end{equation}
where $\nu_{m}$ is the magnetic diffusivity. We consider equation 
(\ref{inde}) in the very large magnetic Reynolds number 
($\propto 1/\nu_m$) limit, which is the case for an accretion disk.
We further have to supply the energy balance equations for ions and electrons by taking into account the 
detailed balance of heating, cooling and advection. 
The magnetized energy equations for ions and electrons read as
\begin{equation}
\Gamma_{3} ' \left[ v_{r}\left\lbrace\frac{\partial p}{\partial r}-\Gamma_{1}\frac{p}{\rho}\frac{\partial \rho}{\partial r}\right\rbrace+v_{z}\left\lbrace\frac{\partial p}{\partial z}-\Gamma_{1}\frac{p}{\rho}\frac{\partial \rho}{\partial z}\right\rbrace \right] 
=Q^{+}-Q^{ie}, 
\label{eq:energy1}
\end{equation}
where
\begin{equation*}
	\Gamma_{1}=\frac{32-24\beta -3\beta^{2}+\frac{2\beta (4-3\beta)}{3\beta_{M}}}{24-21\beta} \ , \ \text{and} \ \Gamma_{3} '=\frac{24-21\beta}{2(4-3\beta)},
\end{equation*}
\begin{equation}
\Gamma_{3} ' \left[ v_{r}\left\lbrace\frac{\partial p_{e}}{\partial r}-\Gamma_{1}\frac{p_{e}}{\rho}\frac{\partial \rho}{\partial r}\right\rbrace+v_{z}\left\lbrace\frac{\partial p_{e}}{\partial z}-\Gamma_{1}\frac{p_{e}}{\rho}\frac{\partial \rho}{\partial z}\right\rbrace \right] 
=Q^{ie}-Q^{-}, 
\label{eq:energy2}
\end{equation}
where $Q^+$ represents the viscous and magnetic (Ohmic) heats generated in the flow, $Q^{ie}$ the Coulomb coupling estimating
the amount of heat transferred from ions to electrons, and finally $Q^{-}$ the radiative cooling rate through electrons via different cooling processes including bremsstrahlung, synchrotron and inverse Comptonization of soft photons supplied from the Keplerian disk.
Various cooling formalisms are adopted from past works \cite{ny95,mc05,rm10}.
In order to solve the equations semi-analytically, we make a reasonable hypothesis in the disk-outflow symbiotic region that
the vertical variation of any dynamical variable (say, $A$) is much less than that with radial variation, 
that allows us to introduce $\partial A/\partial z\approx sA/z$, where $s$ is just the degree of scaling and
is a small number. 

\section{Disk hydromagnetics and energetics}

\begin{figure}[h]
\begin{center}
\includegraphics[width=5.5in]{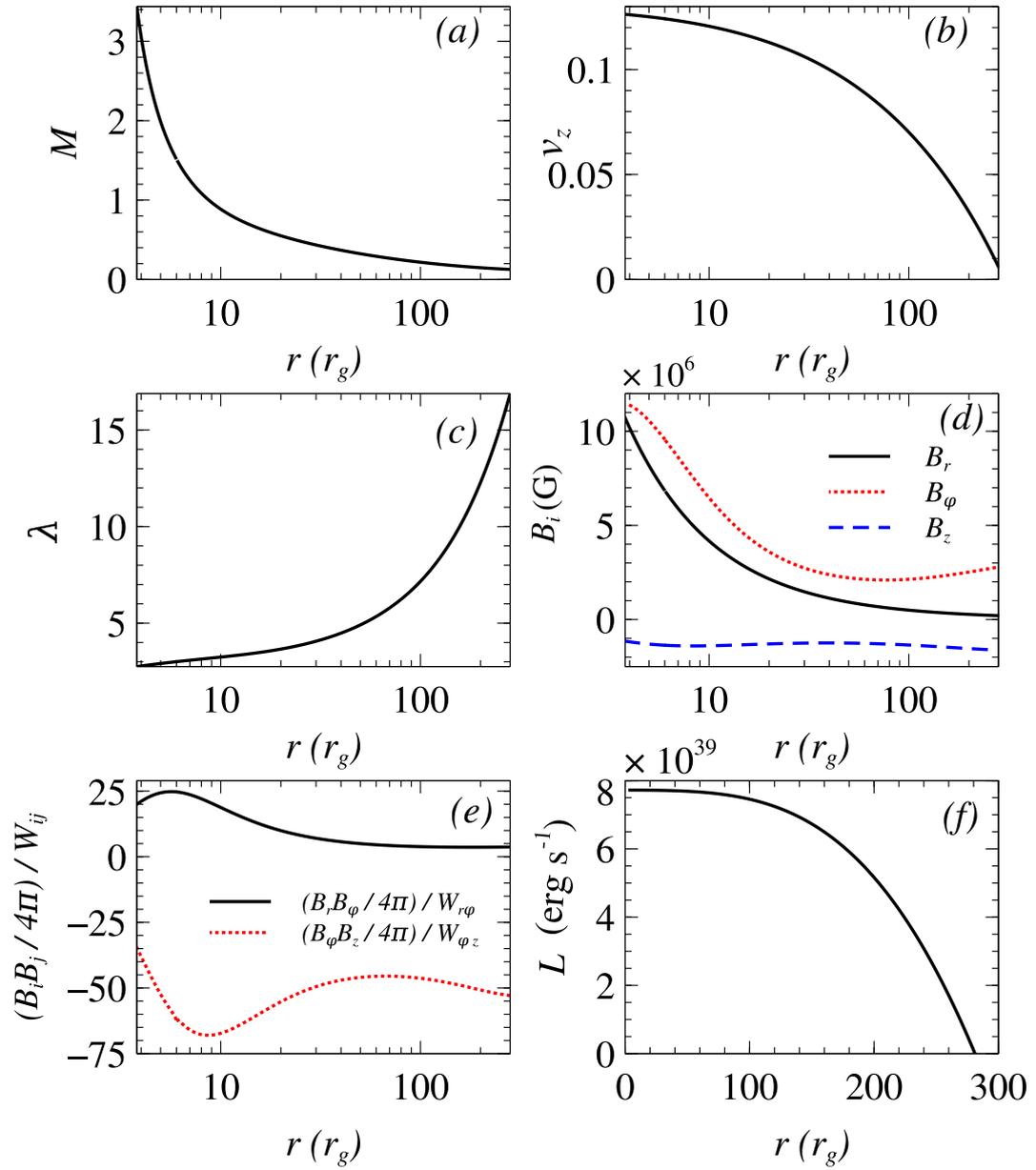}
\end{center}
\caption{Variation of (a) Mach number, (b) outflow speed, (c) specific angular momentum, (d) magnetic field components,
(e) ratio of magnetic to viscous shearing stresses, and (f) luminosity, as functions of distance from the black hole.
Other parameters are $M_{BH}=20M_\odot$, $\dot{m}=0.05$ Eddington rate.}
\label{fig:hydmag}
\end{figure}

Figure \ref{fig:hydmag} shows disk-outflow hydromagnetics revealing that large scale strong magnetic fields
are able to transport angular momentum adequately rendering further significant $v_r$ and $v_z$ with 
decreasing $r$. The angular momentum profile turns out to be similar to that obtained based purely on
$\alpha-$viscosity and hence $W_{ij}$ when the field is weak. However, the benefit with large scale magnetic stress is that it 
renders significant vertical outflow along with radial inflow. It is confirmed from Figs. \ref{fig:hydmag}d,e that close
to the black hole magnetic field could be even $\sim 10^7$ G with an efficient magnetic shear compared to $\alpha-$viscosity 
induced viscous shear. 

Now energetics can be estimated based on above hydromagnetism. 
The energy equation in conservative form under steady state condition is given by
\begin{equation}
\nabla.\mathbf{\mathcal{F}}=0\,\,{\rm with}\,\,
\mathcal{F}_{i}=\rho v_{i} \left(  \frac{v^{2}}{2}+\frac{\Gamma_{1}}{\Gamma_{1} -1} \frac{p}{\rho}+\frac{B^{2}}{8\pi}+\Phi \right)+v_{j}M_{ij}-v_{j}W_{ij},
\end{equation}
where $i,j$ correspond to $r$ or $\phi$ or $z$, indicating radial or azimuthal or vertical component of the respective variables
with $v^{2}=v_{r}^{2}+\lambda^{2}/r^{2}+v_{z}^{2}$, $\Phi$ is the gravitational potential, and $M_{ij}$ is the magnetic stress tensor with standard definition, given by
\begin{equation}
M_{ij}=\frac{B^{2}}{8\pi}\delta_{ij}-\frac{B_{i}B_{j}}{4\pi}.
\end{equation}
The outflow power extracted from the disk is computed at the disk-outflow surface region. It defines as \cite{mm19}
\begin{multline}
        P_{j}(r)=\int 4\pi r  \Bigg[ \rho v_{z} \left\lbrace  \frac{v^{2}}{2}+\frac{\Gamma_{1}}{\Gamma_{1} -1}     \frac{p}{\rho}+\Phi - \Bigg(\frac{\lambda}{r}W_{\phi z}+v_{r}W_{rz}\Bigg) \right\rbrace  \\
        +\frac{v_{z}}{4\pi} \Bigg( B_{r}^{2}+B_{\phi}^{2}-\frac{v_{r}}{v_{z}} B_{r} B_{z} -\frac{\lambda}{rv_{z}} B_{\phi} B_{z} \Bigg) \Bigg]_{h} dr.
\end{multline}
This accretion induced outflow power contains contributions from mechanical and enthalphy powers, and those of viscous and Poynting parts. Our model is restricted vertically up to the disk-outflow 
coupled region, above which outflow may decouple and accelerate. Hence this computed power is basically the initial power of any astrophysical jets at the launching region. Also the disk luminosity can be computed from the cooling mechanisms and can be defined as
\begin{equation}
L=\int \left(\int_{0}^{h} Q^{-}4\pi r \ dz \right)\ dr.
\end{equation}
The variation of disk luminosity, whose magnitude is the most important observable in the present context, is shown in 
Fig.~\ref{fig:hydmag}f. At an arbitrary $r$, the luminosity is obtained by integrating from the outer disk radius $r_{out}$ to that corresponding $r$. 
For the case of a stellar-mass black hole of mass $M_{BH}=20 \ M_{\odot}$ with total mass accretion rate 
$\dot{m}=0.05$ Eddington rate, the maximum attainable luminosity, based on the integration over whole disk, is $L\sim 8\times 10^{39}$ erg s$^{-1}$. 
This value is quite adequate to explain observed luminosities of ULXs in hard states. Table~1 enlists some ULXs with their respective
power-law indices, indicating their harder nature. It is very
interesting that the luminosity of the sources $L\sim 10^{40}$ erg s$^{-1}$, which can be explained by a stellar mass black hole 
accreting at a sub-Eddington rate in the presence of strong magnetic fields, as described in Fig. ~\ref{fig:hydmag}f.

\begin{table}
        \centering
        {Table~1: Some ULX sources in a hard power-law dominated state.}
        \label{tab:table1}
                \begin{tabular}{lccr} 
                        \hline
                        Source & $\Gamma$ & $L_{0.3-10 \ \text{keV}}$  \\
                        &          & $(10^{40}$ erg  s$^{-1} )$  \\
                        \hline
                        M99 X1 \cite{ulx2} & $1.7^{+0.1}_{-0.1}$ & 1.9 \\
                        \hline
                        Antennae X-11 \cite{ulx3} & $1.76^{+0.05}_{-0.05}$ & 2.11 \\
                        \hline
                        Holmberg IX X-1 \cite{ulx4} & $1.9^{+0.1}_{-0.02}$ & 1.0 \\
                        \hline
                        NGC 1365 X1 \cite{ulx5} & $1.74^{+0.12}_{-0.11}$ & 2.8 \\
                        NGC 1365 X2 \cite{ulx5} & $1.23^{+0.25}_{-0.19}$ & 3.7  \\
                        \hline
                        M82 X42.3+59 \cite{ulx6} & $1.44^{+0.09}_{-0.09}$ & 1.13 \\
                        \hline
                \end{tabular}
\end{table}






\section{Summary}
ULXs and the question of plausible existence of intermediate mass black hole in the universe are both puzzling 
us for quite sometime. Some authors argue ULXs to be the sources of an intermediate mass black hole. 
Some others argue ULXs to be super-Eddington acrretors by a stellar mass black hole. The later group further argues
that there is no need to expect a continuous mass distribution of black holes from stellar mass to
supermassive scales. We suggest quite differently and uniquely. We show that at least some of 
ULXs are nothing but the highly magnetized accreting sources of stellar mass black holes accreting at 
a sub-Eddington rate only. The required field magnitude is of the order of $10^7$ G to explain ULXs in hard
states, which is well below the underlying Eddington value. Therefore, at least some of ULXs could 
just be stellar mass black holes. While this suggestion leaves the question for the existence of intermediate
mass black hole wide open, it argues for the power of magnetically dominated/arrested accretion flows
to explain enigmatic astrophysical sources.

\section*{Acknowledgment}
The author thanks Tushar Mondal of IISc for discussion and drawing the figure.

\end{document}